\documentclass{article}

\usepackage{arxiv}

\usepackage[utf8]{inputenc} 
\usepackage[T1]{fontenc}    
\usepackage{hyperref}       
\usepackage{url}            
\usepackage{booktabs}       
\usepackage{amsfonts}       
\usepackage{nicefrac}       
\usepackage{microtype}      
\usepackage{lipsum}
\usepackage[english]{babel}

\usepackage{graphicx}
\usepackage{microtype}
\usepackage[nobreak]{cite}
\usepackage{amsmath}
\usepackage{amsthm}
\usepackage{amsfonts}
\usepackage{amssymb}
\usepackage{graphicx}
\usepackage{wrapfig}
\usepackage[ruled,vlined]{algorithm2e}
\usepackage[dvipsnames, table]{xcolor}
\usepackage{multicol, multirow, booktabs}
\usepackage{xfrac} 
\usepackage{todonotes}
\usepackage{siunitx}
\usepackage{array,ragged2e}

\theoremstyle{definition}
\newtheorem{definition}{Definition}[section]

\SetKwRepeat{Do}{do}{while}

\newcommand{\species}[1]{\ensuremath{\mathtt{#1}}}

\newcommand{\var}[1]{\ensuremath{\mathtt{#1}}}
\newcommand{\vars}{\ensuremath{\mathcal{V}}}
\newcommand{\params}{\ensuremath{\mathcal{P}}}

\newcommand{\statesOf}[1]{\ensuremath{\mathrm{\Pi}(#1)}}
\newcommand{\asyncOf}[1]{\ensuremath{Async(#1)}}
\newcommand{\BN}{\ensuremath{\mathcal{B}}}
\newcommand{\ParBN}{\ensuremath{\widehat{\mathcal{B}}}}

\newcommand{\attractors}{\ensuremath{\mathbb{A}}}
\newcommand{\SB}{\ensuremath{\mathbb{SB}}}
\newcommand{\WB}{\ensuremath{\mathbb{WB}}}

\newcommand{\context}[1]{\ensuremath{\mathcal{T}(\mathtt{#1})}}
\newcommand{\update}[1]{\ensuremath{F_{\mathtt{#1}}}}

\newcommand{\sub}[3]{#1[#2 \mapsto #3]} 

\newcommand{\pA}{\ensuremath{{\color{BrickRed}\blacklozenge}}}
\newcommand{\pB}{\ensuremath{{\color{LimeGreen}\blacktriangle}}}
\newcommand{\pC}{\ensuremath{{\color{MidnightBlue}\spadesuit}}}
\newcommand{\pD}{\ensuremath{{\color{BlueGreen}\clubsuit}}}

\newcommand*{\thead}[1]{\multicolumn{1}{|c|}{\bfseries #1}}

\newcolumntype{L}[1]{>{\raggedright\let\newline\\\arraybackslash\hspace{0pt}}m{#1}}
\newcolumntype{C}[1]{>{\centering\let\newline\\\arraybackslash\hspace{0pt}}m{#1}}
\newcolumntype{R}[1]{>{\raggedleft\let\newline\\\arraybackslash\hspace{0pt}}m{#1}}

\newcommand{\PBN}[1]{ParBN#1}

\title{Parallel One-Step Control of Parametrised Boolean Networks}

\author{
  Luboš Brim\\
  Masaryk University\\
  \texttt{brim@fi.muni.cz} \\
   \And
  Samuel Pastva\\
  Masaryk University\\
  \texttt{xpastva@fi.muni.cz} \\
  \AND
  David Šafránek\\
  Masaryk University\\
  \texttt{safranek@fi.muni.cz} \\
  \And
  Eva Šmijáková\\
  Masaryk University\\
  \texttt{xsmijak1@fi.muni.cz} \\
}

\begin{document}

\maketitle

\begin{abstract}

Boolean network (BN) is a simple model widely used to study complex dynamic behaviour of biological systems. Nonetheless, it might be difficult to gather enough data to precisely capture the behavior of a biological system into a set of Boolean functions. These issues can be dealt with to some extent using parametrised Boolean networks (\PBN{s}), as it allows to leave some update functions unspecified. In this paper, we attack the control problem for \PBN{s} with asynchronous semantics. While there is an extensive work on controlling BNs without parameters, the problem of control for \PBN{s} has not been in fact addressed yet. The goal of control is to ensure the stabilisation of a system in a given state using as few interventions as possible. There are many ways to control BN dynamics. Here, we consider the one-step approach in which the system is instantaneously perturbed out of its actual state. A naïve approach to handle control of \PBN{s} is using parameter scan and solve the control problem for each  parameter valuation separately using known techniques for non-parametrised BNs. This approach is however highly inefficient as the parameter space of \PBN{s} grows doubly-exponentially in the worst case. In this paper, we propose a novel semi-symbolic algorithm for the one-step control problem of \PBN{s}, that builds on a symbolic data structures to avoid scanning individual parameters. We evaluate the performance of our approach on real biological models.\end{abstract}


\section{Introduction}

Cell reprogramming is currently one of the most critical challenges in
computational biology. The goal of cell reprogramming is to \emph{control} a
cell's phenotype. This ability opens many opportunities, mainly in
regenerative medicine. In order to reach the desired phenotype, the
correct transcription factors must be identified. That is close to
impossible to be done only using \emph{in vitro} biological
experiments due to the very high number of possibilities how the cell
might be interfered with. This is where \emph{in silico} analysis and
computational models of cell dynamics come into play. Formal methods and their integration provide a promising technology that allows fully automatic identification of control strategies by using computational models. 

A cell can be viewed as a set of genes and their mutual
regulators. The compact abstraction of these relationships can be
modelled using \emph{Boolean networks} (BNs). BNs are becoming very
popular means for in silico experiments, as they are both simple and
expressive~\cite{SCHWAB2020571}. Moreover, BNs have applications not
only in molecular biology, but also in many other areas including
circuit theory and computer science. BNs are composed of two essential
parts. The first part is a finite set of Boolean \emph{variables} representing genes or other biochemical substances. The
second part is a set of Boolean \emph{update functions} which specify
the way variables dynamically change their value based on influences
from other variables. Typically, influences among variables are
visualised in the form of a so-called regulatory network displaying the structure
of a BN (an example is shown in Fig.~\ref{fig:example_prn}a).

In BN models, time is considered to be \emph{discrete}. At each time
step, some variables are selected for an update. The scheduling of
updates has a strong influence on the reachable configurations of the
variables. There are two dominant updating paradigms. The
\emph{synchronous} paradigm updates all variables simultaneously, and
thus generates deterministic dynamics. In contrast, the (fully)
\emph{asynchronous} paradigm updates a single non-deterministically selected variable at each time step. In this work, we consider the asynchronous update schedule as it often captures the behaviour of biological systems more realistically compared to its synchronous counter-part~\cite{Zheng2013}. 

At any particular time moment, the tuple of all variables' values in
the BN is called a \emph{state} of the network. The
\emph{state-transition graph} (STG) has as its \emph{nodes} the states, and
each directed edge represents a possible transition from one state to
another one in a single update. The size of the STG grows exponentially with
the number of variables, which causes the state-space explosion
problem. Since the BN is a finite-state system, the state of the system will
in a long-run evolve into a single state (steady-state) or a set of
recurring states (a complex attractor). These steady states or
recurring states are collectively called \emph{attractors} and
correspond with the \emph{terminal strongly connected components} (TSCC) in the STG.

A significant shortcoming of BNs when used for modelling a real phenomena resides in the
necessity to fully specify the update functions. In practice, it is
often complicated to exactly identify Boolean functions from
biological data. However, there is typically a good evidence of the
fact that a variable regulates another one. \emph{Parametrised Boolean
  networks} (\PBN)~\cite{Zou2013,Benes2019} address the possibility to
specify BNs without the precise knowledge of some update functions;
the unknown part is represented in terms of \emph{logical
  parameters}. In \PBN{}s, only regulators of a variable need to be
specified. This allows to capture multiple variants of the possible
actual behaviour of variables without conducting many more expensive
experimental observations. A disadvantage of \PBN{}s is that their analysis is significantly more difficult compared to the non-parametrised case as the edges in the STG change according to the chosen \emph{parametrisations}, i.e., a particular setting of logical parameters in the specification of update functions.

The goal of the \emph{Boolean network control} is to influence the behaviour of the
network so that it stabilises in a particular attractor. A typical way
to change the behaviour is to perturb the values of some variables. In
this paper, we consider \emph{one-step state perturbations} in which
we apply all the perturbations of variables simultaneously and only
once. After the perturbation, the system is left to behave
normally. Solutions of the Boolean network control problem for a
model of a cell \emph{in silico} provide the basis for experimental
designs allowing to reprogram the cell \emph{in vitro}. Since
practical realisation of particular perturbations requires
non-trivial effort, the number of perturbations typically needs to be
minimised. To that end, the control problem is usually enriched with some
optimisation criteria. 

In this paper, we focus on the  \emph{source-target} variant of the \PBN{}
control problem where both the source and the target states are given in advance. To
the best of our knowledge, we provide the first efficient solution to
the one-step target control of \PBN{s}. It is worth noting that the parameters bring many new
challenges to the control problem. First, the attractors might significantly
change with the change in the parametrisation of the network. Second, the minimal
state perturbation does not necessarily work for all parametrisations, and
therefore, the notion of the optimal control strategy needs to be adapted to
such a situation. Third, the parameter-space explosion (in addition to
the state-space explosion) makes the problem computationally
demanding. The parameter space is in the worst case, doubly-exponential~\cite{wang2012}. Each \PBN{} parametrisation
generates a unique BN model with a unique STG. That is why using
algorithms developed for asynchronous non-parametrised BNs and
computing the control set for each parametrisation distinctly
(parameter scan) is not feasible, and a new approach needs to be developed.

\subsection*{Contribution}

We propose an efficient alternative to the naïve parameter scan
approach to compute the one-step target control of
ParBNs. Technically, our approach relies on the integration of several formal methods. In particular, we employ a representation of \PBN{s}
based on a~symbolic edge-coloured graph using BDDs~\cite{Bryant86,
	Benes2019}. For this representation, we extend a well-established
algorithm for asynchronous BNs control employing one-step
perturbations~\cite{Baudin2019}. The technique is based on the identification of attractor's strong basin. Our novel algorithm is able to compute the strong basins of all \PBN{} parametrisations simultaneously instead of computing them one by one (individually for every parametrisation). We show that for highly parametrised models, our approach is significantly more efficient than the na\"ive approach.

\subsection*{Related Work}

Typically, two elementary types of control problems are distinguished: (i)
achieving a single
desired target attractor irrespective of the current state (target
control)~\cite{Kim2013}, (ii) achieving control between every pair of attractors (full control)~\cite{fiedler2013dynamics}. Here, we focus on the target control.

The target control problem has been studied in non-parametrised BNs
with both the synchronous and asynchronous update. In synchronous case,
the method~\cite{Kim2013} identifies the control kernel, a
minimal set of nodes, inferred from the explicit
STG of the BN. In~\cite{zhao2015control} a network graph aggregation
approach is employed at the level of the regulatory network avoiding
construction of the full STG. In~\cite{Moradi2019}, the regulatory network is divided to
partially-dependent parts identified as strongly connected components. In
the case of asynchronous BNs, the approach in~\cite{Zanudo2015}
computes a set of relevant BN variables based on the identification of
particular motifs in the regulatory network.

Traditional approaches mentioned above assume the control
to be implemented by involving a permanent perturbation applied
continuously for an extended period of time. However, this is not possible to be
efficiently achieved in practise~\cite{cornelius2013realistic}. To
that end, the concept of one-step perturbation has been
introduced in~\cite{Baudin2019} in the context of asynchronous BNs (the
perturbation is applied in the initial state, and after that
 the system evolves according to its original
dynamics). The algorithms in~\cite{Baudin2019} address both target and
full control, and they solve the existential problem as well as the
optimal problem (identifying the minimal set of perturbations). The method is based on an efficient identification of
a strong basin. In this paper, we employ a similar idea for the
target control in the novel context of \PBN{s}. 

It is also worth noting that in~\cite{MandonHP17, JunPang2020a} the authors work with
the concept of temporary perturbations filling the gap
between one-step and permanent perturbations. Moreover, complex control
strategies considering temporal sequences of perturbations are studied
in~\cite{Jaoude16,Mandon2019CMSB,Pardo19,JunPang2020b}. All those
results are developed for non-parametrised BNs only. 

The approaches mentioned in this subsection cannot be
directly lifted to work with \PBN{s} (the na\"ive solution considering
iteration of existing methods is infeasible due to the combinatorial
explosion of the parameter space). As already stated above, we propose the first efficient solution to the
one-step target control of \PBN{s}. 

\section{Preliminaries}

In this section, we introduce Boolean networks (BNs) and define related
terms regarding long-term behaviour of BNs. Then we expand the notion of BNs by adding parameters, allowing for unspecified or unknown behaviour in the network.

\subsection{Boolean Networks}

Boolean networks are a simple model widely used to study complex dynamic behaviour of biological systems. That is why we define Boolean networks in a way that closely relates to \emph{regulatory networks}, which represent biological processes using directed dependency graphs of biochemical entities:

\begin{definition}[Boolean network]
    \label{def:BN}
    A~\emph{Boolean  network} is a tuple $\BN = (\vars, R, \mathcal{F})$ such that:
    
    \begin{itemize}
        \item $\vars = \{ \var{A}, \var{B}, \ldots \}$ is a finite set of Boolean \emph{state variables}.
        \item $R \subseteq \vars \times \vars$ is a set of \emph{regulations}. For $\var{A} \in \vars$, we say that $\context{A} = \{ \var{B} \in \vars \mid (\var{B}, \var{A}) \in R \}$ is the \emph{context} of $\var{A}$, i.e. the subset of $\vars$ regulating $\var{A}$.
        \item $\mathcal{F} = \{ \update{A} \mid \var{A} \in \vars \}$ is a family of \emph{logical update functions}. The signature of each $\update{A}$ is given by the  context of $\var{A}$ as $\update{A}: \{0, 1\}^{\context{\var{A}}} \to \{0, 1\}$.
    \end{itemize}
\end{definition}

\subsubsection{State space}

A \emph{state} $s$ of a BN $\BN$ is a valuation of its Boolean variables,
i.e. $s: \vars \rightarrow \{0, 1\}$. The set of all possible states
 is $\statesOf{\BN} = \{0, 1\}^\vars$ and is called the \emph{state space} of $\BN$. Given a
state $s$, $\sub{s}{\var{A}}{b}$ denotes a copy $s$ where the
value of $\var{A}$ is set to $b \in \{ 0, 1 \}$. 
Finally, for a state $s$ and an update function $\update{A}$, we use the abbreviated notation $\update{A}(s)$ to denote $\update{A}$ applied to $s$ restricted to the context of $\var{A}$.

\subsubsection{Hamming difference and distance}

The states of BN are Boolean (binary) configuarations of variables. That is why we can conduct standard Hamming operations on them. Given two states $s, s'$, their \emph{Hamming difference} $\mathrm{hdif}(s, s') $vis the set of all variables in which the two states differ:
$\mathrm{hdif}(s, s') = \{\var{A} \in \vars \mid s(\var{A}) \neq
s'(\var{A})\}$. The \emph{Hamming distance} is then the cardinality of
this set, $\mathrm{hdis}(s, s') = |\mathrm{hdif}(s, s')|$.

\subsubsection{Regulatory network}
With every BN \BN{} it is possible to associate a directed graph
$(\vars, R)$ called a \emph{regulatory network} or a \emph{dependency
  graph} of $\BN$. This graph captures influences among the variables
of $\BN$. When visualising a BN, its regulatory network is usually
displayed as a directed graph (with update functions specified
separately). For a BN \BN{}, one also often considers certain general
properties of its regulations, which can then be depicted
in the regulatory network.

\subsubsection{Regulation types}
We say that a regulation $(\var{A}, \var{B}) \in R$ is \emph{observable} if there exists a state such that changing the value of $\var{A}$ also changes the value of $\update{B}$, formally:
\begin{equation*}
\exists s \in \statesOf{\mathcal{B}} : \update{B}(\sub{s}{\var{A}}{0}) \not= \update{B}(\sub{s}{\var{A}}{1})
\end{equation*}

Intuitively, this means that the presence of one biochemical entity
has an observable influence on another entity. When a regulation is
not marked observable, it can have an influence on the regulated
entity, but we do not enforce it. Such regulations are drawn with a
question mark.

In addition to observability, we also consider two possible
\emph{monotonicity} properties of a regulation: \emph{activation} and
\emph{inhibition}. Regulation is activating if by increasing $\var{A}$
it is not possible to decrease $\update{B}$. For example, if $\var{A}$
and $\var{C}$ activate $\var{B}$, the possible functions $F_B$ are
$\var{A} \lor \var{C}$ or  $\var{A} \land \var{C}$. On the contrary,
regulation is inhibiting, if by increasing $\var{A}$ one can not
increase the value of $\update{B}$. For example, if $\var{A}$ and
$\var{C}$ inhibit $\var{B}$, the possible functions $\update{B}$ are
$\neg \var{A} \lor \neg \var{C}$ or $\neg \var{A} \land \neg
\var{C}$. There can be regulations which are neither activating nor
inhibiting. However, most regulations in this paper are either
inhibiting or activating as this is typical for biological
models. Graphically, activating regulation is depicted as a regular
green arrow while inhibiting regulation is drawn as a flat red arrow
(see Fig.~\ref{fig:example_prn}a).

\subsubsection{Dynamics}
The complete dynamical behaviour of a boolean network $\BN$ is captured by the directed \emph{state-transition graph} (STG) $\mathcal G = (S,T)$, where $S=\statesOf{\mathcal{B}}$ and $T\subseteq S\times S$. The definition of the \emph{transition relation} $T$ depends on the \emph{updating scheme} that defines the way variables update their states along time. 
In this paper, we consider \emph{asynchronous} updating. At a discrete time step, the system non-deterministically applies
some $\update{A} \in \mathcal F$ to a state $s$. We then obtain a \emph{transition relation} $\rightarrow$ which is defined as follows:

$$ (s, t) \in \rightarrow \text{ if and only if } s \neq t \land \exists \update{A} \in \mathcal{F} : \sub{s}{\var{A}}{\update{A}(s)} = t $$

Note that for every $(s, t) \in \rightarrow$ the Hamming distance $\mathrm{hdis}(s,t)=1$ as during one step only one variable can change its value.  For $(s,t) \in \rightarrow$, we simply write $s \to t$. We use $\rightarrow^*$ to denote a reflexive and transitive closure of $\rightarrow$, also writing $s \to^* t$ when $(s, t) \in \to^*$.  Also note that the transition relation is \emph{non-deterministic}. We denote $\asyncOf{\BN} $ the state-transition graph of $\BN$ under asynchronous updating scheme.

When studying the long-term behaviour of a BN, we typically only consider \emph{fair} infinite paths in the state-transition graph. In a fair path, if a transition is enabled infinitely often, it has to be taken infinitely often. Therefore, the system cannot infinitely delay the available transitions, and it is not possible to cycle forever in a non-terminal strongly connected component.

\subsubsection{Attractors}
The long-term behaviour of BNs is captured by the notion of \emph{attractors}. In biological models, we observe a phenotype in which the system eventually stabilises, whereas, in BN computational model, we observe attractors which are understood as terminal strongly connected components of the STG. In the following, we use these two terms interchangeably.

\begin{definition}[Attractor]
    Let $\BN = (\vars, R, \mathcal{F})$ be a BN. An
    \emph{attractor} of $\BN$ is a~terminal strongly connected
    component (TSCC) in $\asyncOf{\BN}$, i.e. a~maximal subset $A \subseteq \statesOf{\BN}$
    such that for all $s,t \in A$, $s \to^* t$, and for all $s \in A$
    and $t \in \statesOf{\BN}$, $s \to t$ implies $t \in A$. 
\end{definition}

We denote
a set of all attractors of $\BN$ as $\attractors(\BN)$ or simply as
$\attractors$ if $\BN$ is clear from the context. For a fixed state
$s$, we define $Att(s)$ to be an attractor $A \in \attractors$ such
that $s \in A$, or an empty set when $s$ does not belong to any
attractor. Furthermore, for an attractor $A \in \attractors(\BN)$  we
also define notion of its weak and strong basins. 

\begin{definition}[Weak and strong basin]
 Let $\BN = (\vars, R, \mathcal{F})$ be a BN and $A$ its attractor. A \emph{weak basin} of $A$ is the set of states from which it is
possible to reach $A$:

$$\WB(\BN, A) = \{ s \in \statesOf{\BN} \mid s \to^* t \text{ for some
} t \in A \} $$

A \emph{strong basin} is a set of states from which it is \emph{not} possible to reach \emph{any other} attractor than attractor $A$: 

$$\SB(\BN, A) = \WB(\BN, A) \setminus \bigcup_{A' \in \attractors [A'
  \not= A]} \WB(\BN, A')$$
\end{definition}

\begin{figure}
    \centering
    
    \includegraphics[width=0.6\textwidth]{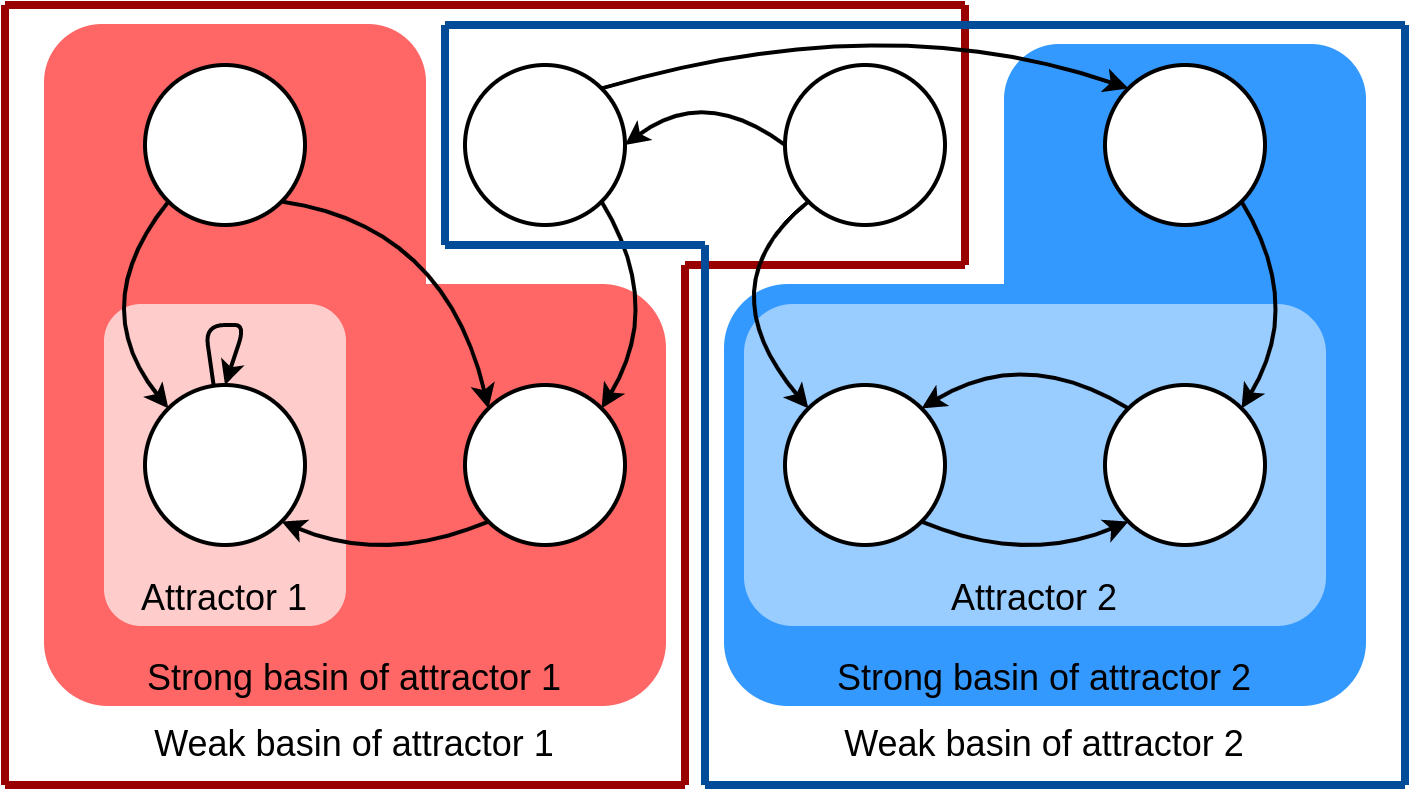}
    \caption{
        Attractors, weak basins and strong basins in a BN. The BN contains two attractors: single-state attractor 1 (light-red area) and cyclic two-states attractor 2 (light-blue area). Both these attractors have strong basins of size 3 (solid red area for attractor 1, blue are for attractor 2 resp.). Note that states of attractors are also parts of their basins. Moreover, the strong basins never have any intersections as given by definition. Finally, the red-lined and blue-lined areas contain weak basin states of attractor 1 and attractor 2. The strong basin is always a subset of a weak basin. The weak basins are over-lapping.  
    }
    \label{fig:basins}
\end{figure}

Notice that due to the fairness property, once a strong basin of
an attractor $A$ is reached, the system eventually stabilises in the
given attractor. For better illustration, Figure~\ref{fig:basins} depicts weak and strong basins of some BN.

\subsection{Parametrised Boolean Networks}


Given a complex real-life system it might be very challenging to precisely determine all the update functions $\mathcal{F}$ of a Boolean network. \emph{Parametrised Boolean networks} \cite{Zou2013,Benes2019} provide a framework to deal with the lack of precise knowledge about the updating mechanism in a system. This extension assumes a set of logical parameters which determine the behaviour of update functions. Therefore, parametrised logical update functions either return a Boolean value (they behave normally) or a logical parameter representing the uncertainty of the consequent behaviour:

\begin{definition}[Parametrised Boolean network]
    We define a \emph{parametrised Boolean network (\PBN{})} to be a tuple $\ParBN = (\vars, \params, R, P, \mathfrak{F})$. Here, $\vars$ and $R$ are the same as in Definition~\ref{def:BN} and
    \begin{itemize}
        \item $\params = \{ \var{P}, \var{Q}, \ldots \}$ is a finite set of Boolean \emph{logical parameters};
        \item $P \subseteq \{ 0,1 \}^{\params}$ is a subset of valid \emph{parametrisations}; 
        \item $\mathfrak{F} = \{\widehat F_\var{A}\mid \var{A} \in \vars\} $ is a family of \emph{parametrised logical update functions}. The signature of each $\widehat F_\var{A}$ is given as
$\widehat F_\var{A} : \{0, 1\}^{\context{A}} \to
(\{0, 1\} \cup \params)$.
    \end{itemize}
\end{definition}

\begin{figure}[t]
    \centering
    \begin{minipage}{0.25\linewidth}
        \centering
        \includegraphics[width=0.9\linewidth]{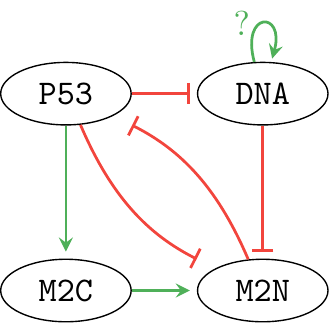}
    \end{minipage}
    \begin{minipage}{0.40\linewidth}
        \begin{center}
            \setlength\tabcolsep{0.5pt} 
            \begin{tabular}{ C{15pt} C{15pt} C{15pt} | C{17pt} | C{9pt} C{9pt} C{9pt} C{9pt} C{9pt} C{9pt} C{9pt} C{9pt} C{9pt}}
                \species{M2C} & \species{DNA} & \species{P53} & $\widehat{F}_{\var{M2N}}$ & \multicolumn{9}{c}{$\update{M2N}$ } \\ \hline
                0 & 0 & 0 & $\var{P}_1$
                & 0 & 0 & 1 & 1 & 1 & 1 & 1 & 1 & 1 \\
                0 & 0 & 1 & $\var{P}_2$
                & 0 & 0 & 0 & 0 & 0 & 0 & 0 & 1 & 1 \\
                0 & 1 & 0 & $\var{P}_3$
                & 0 & 0 & 0 & 0 & 0 & 0 & 1 & 0 & 1 \\
                0 & 1 & 1 & 0
                & 0 & 0 & 0 & 0 & 0 & 0 & 0 & 0 & 0 \\
                \hline
                1 & 0 & 0 & 1
                & 1 & 1 & 1 & 1 & 1 & 1 & 1 & 1 & 1 \\
                1 & 0 & 1 & $\var{P}_4$
                & 0 & 1 & 0 & 1 & 1 & 1 & 1 & 1 & 1 \\
                1 & 1 & 0 & $\var{P}_5$
                & 0 & 1 & 1 & 0 & 1 & 1 & 1 & 1 & 1 \\
                1 & 1 & 1 & $\var{P}_6$
                & 0 & 0 & 0 & 0 & 0 & 1 & 0 & 0 & 1 \\
            \end{tabular}
        \end{center}
    \end{minipage}
    \begin{minipage}{0.33\linewidth}
        \flushright
            \begin{tabular}{ C{15pt} C{15pt} | C{17pt} | C{8pt} C{8pt} C{8pt} }
                \species{DNA} & \species{P53} & $\widehat{F}_{\var{DNA}}$ & \multicolumn{3}{c}{\update{DNA}} \\ \cline{1-6}
                0 & 0 & $\var{P}_7$ & 0 & 1 & 1 \\
                0 & 1 & 0 & 0 & 0 & 0 \\
                1 & 0 & 1 & 1 & 1 & 1 \\
                1 & 1 & $\var{P}_8$ & 0 & 0 & 1 \\
            \end{tabular}
            
            \vspace{5pt}
            \begin{tabular}{ C{15pt} | C{17pt}}
                \var{P53} & \update{M2C} \\ \hline
                0 & 0 \\
                1 & 1 \\            
            \end{tabular}
            \hspace{2pt}            
            \begin{tabular}{ C{15pt} | C{17pt}}
                \var{M2N} & \update{P53} \\ \hline
                0 & 1 \\
                1 & 0 \\            
            \end{tabular}
            \hspace{2pt}    
    \end{minipage}
    
    \vspace{5pt}
    
    \begin{minipage}{0.25\linewidth}
        \centering (a)
    \end{minipage}
    \begin{minipage}{0.40\linewidth}
        \centering (b)
    \end{minipage}
    \begin{minipage}{0.33\linewidth}
        \centering (c)
    \end{minipage}
    
    \caption{ (a) A regulatory network of a simplified \PBN{} describing the
        DNA damage mechanism adapted
                from~\cite{prnExample}. Every regulation is either
                activating (green) or inhibiting (red) and observable,
                except for $(\var{DNA}, \var{DNA})$, which is not
                necessarily observable.
        (b) All possible valid update functions $\update{M2N}$ satisfying the static constraints (monotonicity, observability). (c) Valid update functions $\update{DNA}$, $\update{M2C}$ and $\update{P53}$ satisfying the static constraints. Here, $\var{P}_i$ denote the parameters ($\params$) of the \PBN{}.}
    \label{fig:example_prn}
\end{figure}

For $p \in P$, we write $p(\var{P})$ to denote the value of $\var{P}$ in $p$ and we also use the same notation $\sub{p}{\var{P}}{k}$ for substitution as we used for states. The notion of the state space of a \PBN{} is identical to that of a BN. By fixing $p \in P$, we obtain  $\ParBN_p = (V, R, \mathfrak{F}(p))$ (a standard BN), $\attractors_p$ (the set of attractors of $\ParBN_p$), and $Att_p(s)$  (the attractor of state $s$ in the parametrisation $p$). 

\begin{figure}
    \centering
    
    \includegraphics[width=0.6\textwidth]{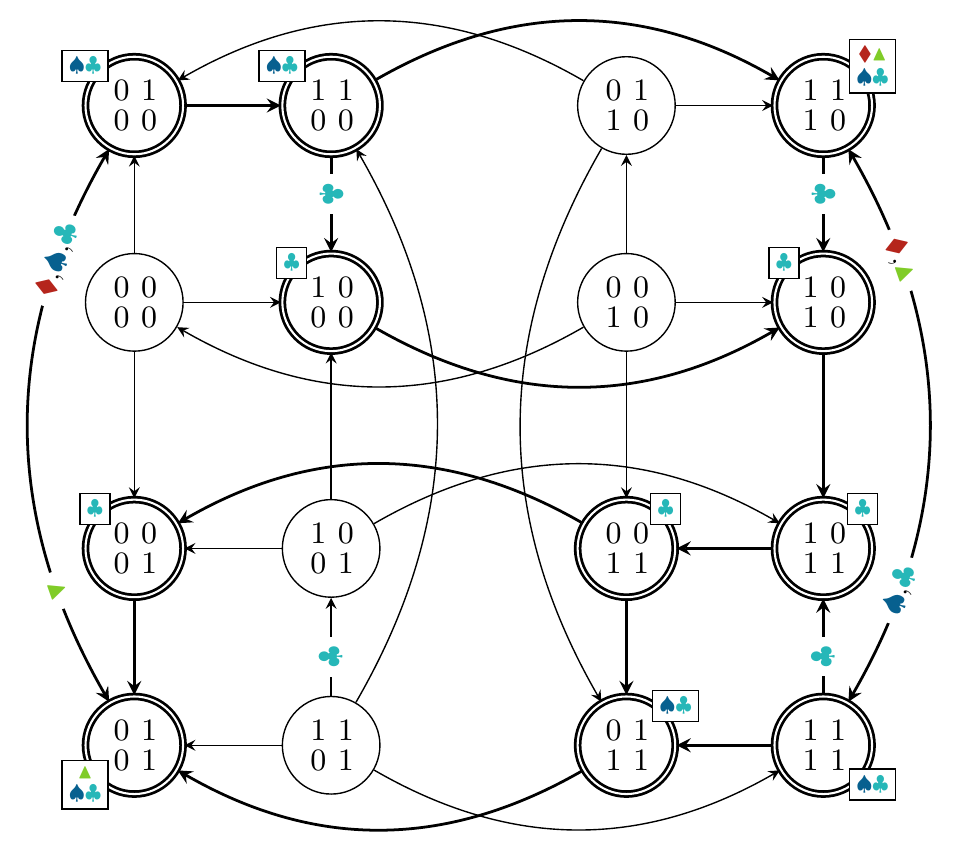}
    \caption{
        The asynchronous semantics of the \PBN{} given in Fig.~\ref{fig:example_prn}a,
        restricted to $P = \{ \pA, \pB, \pC, \pD \}$. Here, $\pA = \{ \var{P}_{2,3,6}: 0, \var{P}_{1,4,5,7,8}: 1 \}$, $\pB = \sub{\pA}{\var{P}_3}{1}$, $\pC = \sub{\pA}{\var{P}_6}{1}$, and $\pD = \sub{\pC}{\var{P}_8}{0}$. The unlabelled edges are enabled for all parametrisations. The highlighted vertices represent attractors for indicated parametrisations.
    }
    \label{fig:example_graph}
\end{figure}

\subsubsection{Dynamics}
The asynchronous semantics of a \PBN{} $\ParBN$ is represented using
an \emph{edge-labelled state-transition graph} $\asyncOf{\ParBN}$,
where each transition $s \to t$ is labelled with a subset of
parametrisations $P(s, t) \subseteq P$ for which it is enabled. That
is, $p \in P(s, t)$ if and only if $s \to t$ in
$\asyncOf{\ParBN_p}$.  For a fixed
$s$, we denote $\mathit{successors}(s)$ the set of all successors of
$s$ (states with Hamming distance one). In Fig.~\ref{fig:example_prn},
we show a small example of a \PBN{}. Fig.~\ref{fig:example_graph} then presents its
asynchronous semantics for selected subset of parametrisations. 

In general, the size of the set of all possible parametrisations
(\emph{parameter space}) can be even doubly-exponential in the number
of Boolean variables. In particular,
the number of Boolean functions in a model with $n$ variables is $2^{2^n}$. It is thus
critical to restrict the parameter space as much as
possible. In many biological models, regulations are usually supplemented with static constraints limiting their outcomes~\cite{Klarner15,streck2016}.

We already presented observability, activation and inhibition as specific properties of regulations. In a parametrised setting, these properties can be used as constraints to restrict the parametrisation space. We assume that every regulation in a \PBN{} can be marked with a subset of these three constraints. Then for all $p \in P$ of $\ParBN$, $\ParBN_p$ must adhere to these constraints, e.g. a regulation marked observable in $\ParBN$ must be observable in $\ParBN_p$ and the same for activation and inhibition.

In Fig.~\ref{fig:example_prn}a, a \PBN{} is displayed where all regulations are marked as either activating or inhibiting. Figures~\ref{fig:example_prn}b and~\ref{fig:example_prn}c then show the possible update functions satisfying these static constraints together with the corresponding logical parameters. Note that the fully parametrised model would have 16 parameters and $65536$ parametrisations, but by applying the static constraints, only $27$ parametrisations remain valid, significantly reducing the size of the associated edge-coloured graph.

\subsubsection{Attractors}
Notice that the standard notion of an attractor cannot be directly transferred to \PBN{s}, because a state can belong to an attractor only in certain parametrisations (for different parametrisations, the attractors do not have to overlap). We say that a subset $A \subset \statesOf{\ParBN}$ is an \emph{attractor in a parametrisation} $p \in P$ if $A$ is an attractor of $\ParBN_p$. Furthermore, given a state $s$, we can define $Ap(s)$ as the subset of $P$, such that for each $p \in Ap(s)$, $Att_p(s) \not= \emptyset$.

\section{Control Problem for Parametrised Boolean Networks}
A computational model is controllable if we can assure that from some
initial state, it reaches a desired final state in a finite amount of
steps. This property was well-studied in the context of synchronous
BNs and is also being pioneered for asynchronous BNs. However, the
control problem for \PBN{s} has not been explored yet. \PBN{s} 
offer a more flexible representation of biological systems by allowing
some uncertainty in the specification of the logic behind
regulations. Since in reality information on regulatory mechanisms is
typically ambiguous or unknown, studying the control problem for
\PBN{s} enables new attractive, real-life applications, such as the discovery of candidate transcription factors for cell reprogramming (i.e., to change a cell's phenotype) when the model is not fully known.

\subsection{Control Problem for Boolean Networks} \label{ss:BNControl}

There are multiple ways to control the behaviour of a BN, mainly differentiated between \emph{state} and \emph{function} perturbations. State perturbations force the system to change its current state. Alternatively, function perturbations adjust the update function(s) and therefore modify the edges of the system's state-transition graph. Typically, changing the processes in a cell (function perturbation) is more difficult then artificially adding or extracting a biochemical substance (state perturbation). For this reason, this paper focuses on state perturbations.

State perturbations can be further differentiated based on their temporal characteristics, mainly \emph{one-step}, \emph{sequential}, \emph{temporary}, and \emph{permanent} control. In one-step control, we change the values of the controlled variables once, and then the network evolves as originally defined. Sequential control identifies a sequence of perturbations that are applied at different time steps. When applying the control temporarily, there exists a finite amount of computational steps after which the control is released. Finally, the most intrusive control is permanent control of variables. However, this scenario is rather unrealistic in practice, as the given substance would need to be added or extracted forever. Moreover, the permanent perturbation might disrupt the original long-term behaviour of the network or introduce completely new behaviours.

Finally, we consider different control objectives, defined in~\cite{Baudin2019} as follows:

\begin{enumerate}
\item \emph{Source-target control}: Given a \emph{source state} $s \in \statesOf{\BN}$, and a \emph{target attractor} $T \in \attractors$, find such control that when applied, the BN always converges from the state $s$ to the attractor $T$.
\item \emph{Target control}: Given a \emph{target attractor} $T \in \attractors$, find a control for every \emph{source attractor} $S \in \attractors$ (such that $S \not= T$) that when applied, the BN converges from $S$ to $T$.
\item \emph{Full control}: For all pairs of distinct attractors $S, T \in \attractors$, find a control which guarantees that the BN converges from $S$ to $T$.
\item \emph{All-pairs control}: Given a subset of source attractors $\mathcal{S} \subseteq \attractors$ and target attractors $\mathcal{T} \subseteq \attractors$, for every pair $S \in \mathcal{S}$ and $T \in \mathcal{T}$, find a control which guarantees that the BN converges from $S$ to $T$.
\end{enumerate} 

In some cases, one may also choose when the control is applied. Here, we consider \emph{immediate} control, i.e. the control applied to the given state. Alternatively, during \emph{sequential} control, the perturbations can be applied multiple times at different time points. This way we can sometimes control the system using fewer perturbations. However, this approach brings new issues, such as dealing with the non-determinism of the BN (different perturbations may be necessary in different branches of the non-deterministic network's behaviour). Moreover, we would need to be able to precisely observe current state of the BN currently. These issues can be addressed to some extent using \emph{attractor-based sequential} approach \cite{Mandon2019CMSB}.

In summary, the control problems for BNs differ in the following aspects:

\begin{itemize}
\item \emph{What do we want to control; goal}: What is the initial state of the BN? Where we want to end? Do we want to control only one scenario or multiple scenarios?
\item \emph{What control we apply}: We can either perturb states (once, temporarily, forever) or functions of variables;
\item \emph{When we apply control}: Only once from an initial state in contrast with applying control to an arbitrary state and any amount of times.
\end{itemize}

\subsection{One-step Control Set of Parametrised Boolean Networks}

In this paper, we focus on state perturbations applied immediately in the initial state. This is the elementary way to perturb the system when solving the source-target control problem. The respective notion of one-step control is stated formally in the following definition.

\begin{definition}[One-step control of \PBN{}]
Given a \PBN{} $\ParBN$, a \emph{one-step state perturbation control} $C$ (further referred simply as \emph{control}) is a tuple $ (\mathbf{1}, \mathbf{0}) $ where $\mathbf{1}, \mathbf{0} \subseteq \vars$, $\mathbf{0}$ and $\mathbf{1}$ are mutually disjoint (possibly empty) subsets of variables of $\ParBN$. The set of all possible controls is denoted $\mathbb{C}$. An \emph{application of control} $C$ to a state $s$, denoted $C(s)$, results in a state $s'$ defined as: 

$$
s'(v) = 
\begin{cases}
    1 & v \in \mathbf{1}\\
    0 & v \in \mathbf{0}\\
    s(v) & \text{otherwise}
\end{cases}
$$
The \emph{size of} $C$ is defined as $\text{Size}(C) = |\mathbf{1}| + |\mathbf{0}|$.
\end{definition}

Next we focus on establishing the control problem (when computing the
one-step perturbation control) in \PBN{s}. Conceptually, the parametrised control problem is to some extent
similar to the \emph{parameter synthesis} problem. The goal of
parameter synthesis is to find parametrisations that ensure some
desired behaviour in the network. Such behaviour can also involve
reachability of specific attractors. However, the key distinction
between parameter synthesis problem and control problem is that in
control, one determines perturbations that lead to a particular
objective (with respect to given parametrisations), possibly resulting
in behaviour that is not achievable only by tuning the network
parameters. In this work, we thus treat parameters as unknown
properties of the system rather than components that can be influenced. 

One of the greatest challenges of \PBN{} control is that the
attractors change based on the parametrisation. To that end, it makes
sense to solve the control problem only for parametrisations which
admit the given attractor. The parametrised control problem then
computes a mapping that associates a potential control with the
maximal set
of parametrisations for which the control is applicable (the
so-called controlled parametrisations). Formally, the considered parametrised
control problem is stated in the following definition.

\begin{definition}[Source-target control in \PBN{}]
Given a \PBN{} $\ParBN$, a source state $s$ and a target
  state $t$, find a mapping $Cp: \mathbb{C} \to 2^{Ap(t)}$ which
assigns each control $C \in \mathbb{C}$ a maximal (possibly empty) set
of parametrisations $p$ for which when $C$ is applied, $\ParBN_p$
converges from $s$ to $Att_p(t)$. We refer to $Cp(C)$ as the \emph{control enabling parametrisations} for $C$.
\end{definition}

Intuitively, control enabling parametrisations $Cp(C) = p$ are the parametrisations, for which the target state is achieved by applying $C$ in the non-parametrised case. Notice, that source-target problem in context of \PBN{} is aiming to drive the network into a target \emph{state} of some attractor instead of an attractor itself. This is because parameterisations might contain attractors which are considered similar as all of them contain the given state. Therefore, in all parameterisations $Ap{t}$ the given state $t$ is entered infinitely often by the controlled BN. If this is not the case of some \PBN{} and some parametersiations contain attractors which are out of the interest, the set of parameterisations $AP(t)$ might be replaced with an arbitrary one.

It is worth noting that it is always possible to bring the system
into the given target state by setting the \PBN{'s} variables to the values adequately (with the values of the given target state). We call this control \emph{trivial}. However,
when controlling a system, we typically look for a control which
requires the fewest interventions as possible. Therefore, we want to minimise the number of the
controlled variables and the trivial solution might not be optimal. In a non-parametrised setting, this is typically the only considered optimisation criterion. 

In a \PBN{}, the situation is further complicated due to the dependence on the parameters. To
reach some attractor, it is sufficient to reach its strong basin after
the application of a control. Nonetheless, the strong basin of an
attractor can vary according to the parametrisation, and the control 
thus "works" only for its control enabling parametrisations. We are interested in maximal sets of control enabling parametrisations. To that end,  we consider the notion of robustness that normalizes the number of control enabling parametrisations. 

\begin{definition}[Robustness of control]
Given a \PBN{} $\ParBN$, a target state $t$, a control set $C$ and $Cp(C)$, the \emph{robustness of control} $C$ is defined as the ratio between  the number of control enabling parametrisations and the number of  all relevant parametrisations: $$Rob(C) = \frac{|Cp(C)|}{|Ap(t)|}$$
\end{definition}

Unfortunately, it is not always possible to ensure that some control is both minimal and the most robust. For example, there may be a control set~$C$ small in size, which only works for a small fraction of the parameter space. Consequently, while $C$ is easily applicable, it may be unlikely to work in reality, as the real behaviour of the system can also follow one of the parametrisations which are not controlled by $C$. We discuss this issue in more detail in Section~\ref{sec:results}.

\section{The Algorithmics}

We are now ready to describe our computational framework for solving the one-step state
perturbation control of \PBN{}. We start by introducing our approach for finding strong
basins in a \PBN{}. Then we explain the framework
for exploring \PBN{'s} STG and for manipulation with \PBN{'s}
parametrisations. Finally, we build a concise workflow for \PBN{}
control employing the proposed algorithms. 

\subsection{Semi-symbolic Parametrised Strong Basin Search Algorithm}

We assume the set of parametrisations of a \PBN{} is represented as a
reduced ordered binary decision diagram (BDD)~\cite{Bryant86}. The
decision variables of the BDD are the parameters of the network
($\params$), meaning that every path from the root to a leaf in such a BDD represents a parametrisation of the \PBN{}. Common logical operations on such BDDs (and, or, negation, ...) then correspond to set operations (intersection, union, complement, ...).  Furthermore, static constraints (activation, inhibition, observability) can be formalised using Boolean formulae over $\params$, and we can, therefore, create a BDD which enforces all imposed constraints and represents the set of all valid parametrisations.

Recall that we represent \asyncOf{\ParBN} as an edge-labelled
state-transition graph, where each transition $s \to t$ has an
associated set of parametrisations $P(s, t) \subseteq P$ represented as a BDD. A \emph{parametrised state set} is a mapping $\statesOf{\ParBN} \to 2^{P}$ assigning to each state a set of parametrisations.  Furthermore, we suppose that the state space is represented \emph{explicitly}, meaning that  all operations on states are performed element-wise (typically in parallel).

We consider parametrised reachability procedures — given a
source state $s$ and a parameter set $P$, these procedures compute a maximal
parametrised state set of all forward/backward reachable states
from the source state $s$ (containing all reachable states $t$ where
each $t$ is
associated with a maximal set of parametrisations $P_t$ for which $t$
is reachable from $s$):

\begin{align*}
\mathit{forwardReachability}(s, P) = \{ t\mapsto P_t \mid \forall p \in P_t: p \in P \land s \to^*_p t \}\\
\mathit{backwardReachability}(s, P) = \{ t\mapsto P_t \mid \forall
p \in P_t: p \in P \land t \to^*_p s \}\\
\end{align*}

The chosen representation (explicit state space and symbolic
parametrisations) allows to compute the reachability procedures in parallel.
For the underlying implementation, we are working with
internal libraries of the tool Aeon \cite{aeon} which provides the most of the necessary functionality, including a convenient format for specifying \PBN{s} and parallel reachability procedures.

The key observation for controlling \PBN{s} is that an attractor (by
definition) is always reached from its strong
basin. From all other states, it is possible to reach also some other
attractor(s); therefore, reaching the target attractor is not
guaranteed. When the attractor's strong basin for all parametrisations
is known, we can compute the control mapping $Cp$ from a source state
to the target attractor by considering Hamming differences between the
source and the states of the strong basin. 

Our algorithm is based on the fixed-point approach for strong basin computation in non-parametrised BNs~\cite{Paul2018CMSB}. The premise is that only states from which it is possible to reach the attractor (weak basin) can be part of the strong basin. The weak basin of the attractor is computed using some standard reachability algorithms, for example, BFS. Then states are iteratively removed if it is possible to leave the basin from them (and therefore not reach the given attractor). Finally, if there are no states left to remove, the fixed-point is achieved, and the strong basin is found.

For parametrised control, we first need to determine $Ap(t)$ for the
target state $t$, since in all other parametrisations, $t$ is not a
part of an attractor. This process is described in
Algorithm~\ref{algo:attractor-params}. In
Algorithm~\ref{algo:strong-basin}, we then extend the approach
from~\cite{Paul2018CMSB} onto \PBN{}. For each state we remember under
which parametrisations it is in the basin, starting with the
parametrised weak basin. Then we iterate over the basin's states. For
each state, we consider its successors, and we compute the
parametrisations for which some successor is \emph{not in the strong
basin}. In these parametrisations, it is possible to leave the
basin, and therefore these parametrisations are not part of the
strong basin, and so we remove them for this state. We can process the states in parallel 
and compute for them the parameterisations which are not part of the strong basin. To avoid 
writing of the strong basin structure while it is intensively read from, we store the parameterisation 
which should be removed separately and after it is computed, we update the strong basin structure 
sequentially. The procedure is iterated until nothing more can be removed from the strong basin.

\begin{algorithm}[h]
\SetKwInOut{Input}{Input}
\SetKwInOut{Output}{Output}
\Input{PBN $\ParBN$, target attractor state $\var{target}$}
\Output{Attractor parametrisations $Ap(\var{target})$}
\BlankLine
$\var{fwd} \gets forwardReachability(\var{target}, P)$\;
$\var{bwd} \gets backwardReachability(\var{target}, P)$\;
\tcc{For every state, compute the BDD difference}
$\var{notAttractor} \gets \var{fwd} \setminus \var{bwd}$\;
\Return{$P \setminus \bigcup_{\var{s} \in \statesOf{\ParBN}} \var{notAttractor}(\var{s})$}\;
\caption{Compute attractor parametrisations $Ap(\var{target})$.}
\label{algo:attractor-params}
\end{algorithm}

\begin{algorithm}[h]
\SetKwInOut{Input}{Input}
\SetKwInOut{Output}{Output}
\Input{PBN $\ParBN$, a state $\var{target}$, attractor parametrisations $Ap(\var{target})$}
\Output{parameterised state set $\statesOf{\ParBN} \to
  2^{Ap(\var{target})}$ representing the strong basin}
\BlankLine
  $\var{SB} \leftarrow backwardReachability(\var{target}, Ap(\var{target}))$\;
  $\var{to\_update} \leftarrow \var{s}$ in $\{ \var{s} \in \statesOf{\ParBN} \mid \var{SB}(\var{s}) \not= \emptyset \}$\;
  \Do{$\var{to\_update} \neq \emptyset$}{
    $\var{updated} \gets \emptyset$\;
    $\var{to\_remove} \gets \forall \var{s} \in \statesOf{\ParBN} \rightarrow \emptyset$\;
    \textbf{parallel} \For{$\var{to\_update}$}{  
         \For {$\var{t}$ in $successors(\var{s})$} {
            \tcc{Recompute parametrisations leading outside of basin}
            $\var{to\_remove}(\var{s}) \gets \var{SB}(\var{s}) \cap P(\var{s}, \var{t}) \cap (P \setminus \var{SB}(\var{t}))$\;
        }
        \uIf{$\var{to\_remove}(\var{s}) \neq \emptyset$} {
              $\var{updated} \gets \var{updated} \cup \{s\}$ \;
        }
    }
    $\var{to\_update} \gets \emptyset$\;
    \For {$\var{t}$ in $\var{updated}$} {
            \tcc{Update strong basin}
            $\var{SB}(\var{s}) \gets \var{SB}(\var{s}) \setminus \var{to\_remove}(\var{s})$\;
            \tcc{Only predecessors of updated vars might be updated in the next iteration}
            $\var{to\_update} \leftarrow \var{to\_updated} \cup predecessors(\var{s})$ \;
        }
}
  \Return{\var{SB}}\;
\caption{Compute strong basin for an attractor in parallel.}
\label{algo:strong-basin}
\end{algorithm} 

\subsection{Control Computation Workflow}

\begin{figure}[ht]
    \centering
    \includegraphics[width=\textwidth]{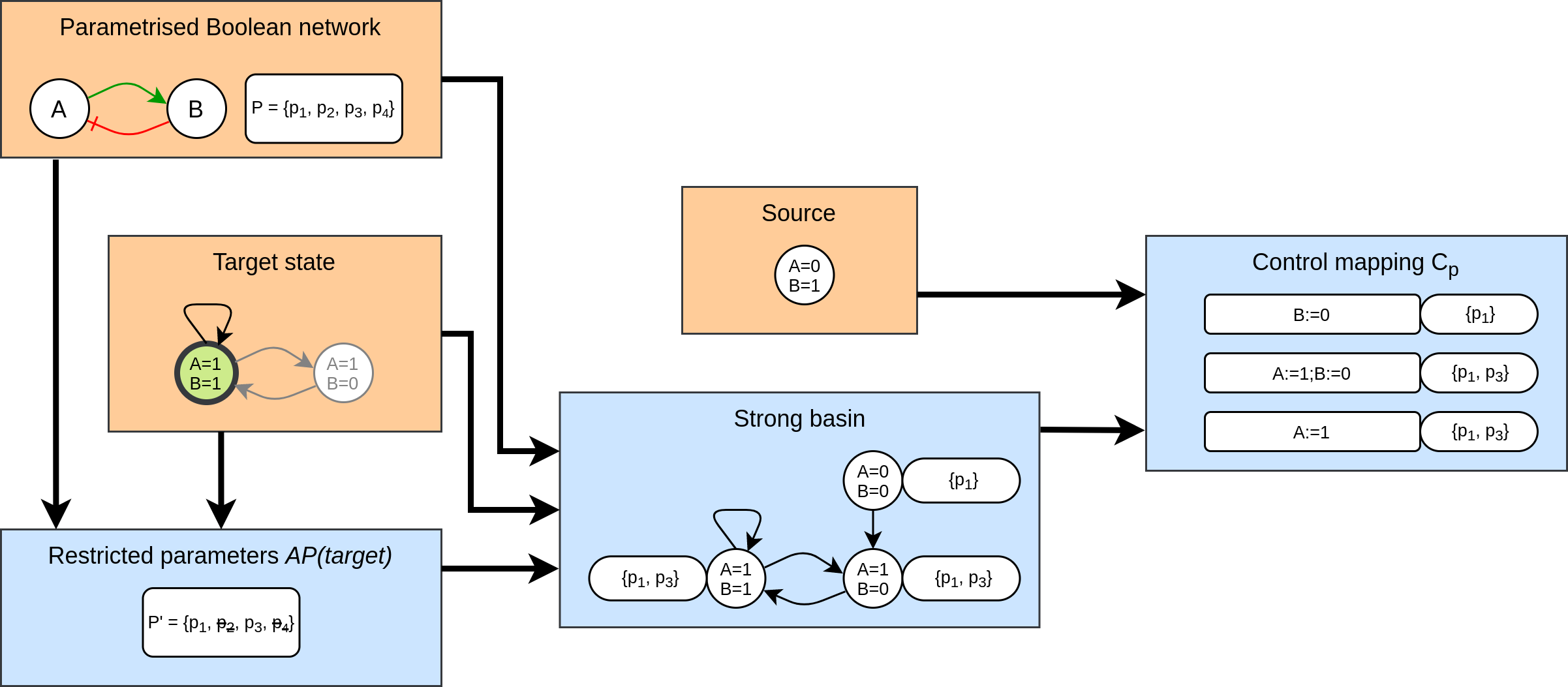}
    \caption{
        Workflow of computing the source-target control problem.
    }
    \label{fig:workflow}
\end{figure}

Now we can describe a complete workflow for computing the source-target control in \PBN{s}, depicted in Fig.~\ref{fig:workflow}. The workflow consists of three inputs and three computation steps, resulting in the control mapping $Cp$. 

Given an input \PBN{} and a target attractor state $target$, we start by computing valid parametrisations set $Ap(target)$ using Algorithm~\ref{algo:attractor-params}. Then, the parametrised strong basin is computed from the \PBN{} with target attractor state $target$ and its valid parametrisations $Ap(target)$ using Algorithm~\ref{algo:strong-basin}. After that, from the strong basin and the source state, we obtain the complete control mapping. Notice that we do not need to know the source state for computing the strong basin and that we can re-use the target's strong basin for obtaining control for different sources. 

To compute the control mapping, observe that viable controls correspond exactly to the Hamming differences (the variables with opposite values) between the source state and the states of the strong basin; yielding one viable control $C$ for every state $s$ in the strong basin $\var{SB}$. Any other control $C'$ does not reach the strong basin and therefore does not guarantee to reach the target attractor (for these, $Cp(C') = \emptyset$). A control $C$ is then viable only for parametrisations for which $s$ appears in the strong basin, we thus set $Cp(C) = \var{SB}(s)$.

Finally, we can compute the size and robustness of each control as we
have all knowledge regarding the variables which need to be controlled
and for which parametrisations the controls work. If it is desired, we
can construct a \emph{witness} BN for a control $C$ (a
non-parametrised BN where the given control works) by fixing some
parameter from $Cp(C)$. The prototype implementation containing all
parts of the workflow is available\footnote{\url{http://github.com/sybila/biodivine-pbn-control}}. 

In highly parametrised models, it is not unusual to obtain a control with size 0, where in some parametrisations no action would be needed to control the model. This is because, in some parametrisations, the source might already be a part of the target's strong basin. If this case is considered unrealistic (since there is probably a need for control, thus these parametrisations appear to be invalid), the parametrisations $Ap(target)$ can be replaced with a custom set of parametrisations.

The resulting set of all available controls can be then used for e.g. cell reprogramming. However, we might obtain many possible controls, and we need to decide which one should be applied. To do that, we can decide based on the size of control or its robustness. The control set can be arbitrarily filtered discarding controls with size bigger than the trivial control (which always has 100\% robustness) or bigger than some set size. Similarly, the control set can be pruned based on too low robustness. The interplay of these two factors is non-trivial, and it is left to the actual application to decide which control would best suit its needs.

\enlargethispage*{5mm}
\section{Evaluation}

\label{sec:results}
We evaluate our approach on two real-life BN models. We compare the performance of our approach using a different number of parameters implanted into the models, resulting in different size of relevant parameter space. We conducted all measurements using a machine equipped with AMD Ryzen Threadripper 2990WX 32-Core Processor and 64GB of memory.

The first model is a \emph{cell-fate decision model}
\cite{Calzone2010}. The model provides a high-level view of possible
different cell fates such as pro-survival, necrosis or apoptosis. We
used a fully-parametrised version (all update functions are completely
parametrised) of this model and we selected seven
biologically relevant attractors. We computed strong basins only for these attractors in several parametrised
versions of the model differing in the number of unknown parameters
(see Table~\ref{table:results}).

The second model, a \emph{myeloid differentiation network}
\cite{Krumsiek2011}, was designed to model a muscle tissue cell
differentiation from common myeloid cell to specialised muscle cells
(megakaryocytes, erythrocytes, granulocytes and monocytes). The
original non-parametrised network has eleven nodes and six
attractors. We derived several parametrised versions of the model by arbitrarily
parametrising update functions of the model. Similarly to the
previous case, we used only attractors of the original network when
computing strong basins for parametrised versions of the model. 
The results are again  shown in Table~\ref{table:results}.

\enlargethispage*{5mm}
\begin{table}[h]
\caption{Results of strong basins computation. The values are stated
  as ranges because we computed strong basins of all attractors
  considered in given models. The second column shows the number of
  model's parameters. The third column shows count ranges of
  parametrisations which contain the given attractor. The fourth
  (resp. fifth) column displays ranges of the number of states in the
  weak (resp. strong) basins. The last column contains ranges of times
  needed to compute strong basins.}
\centering
\setlength{\tabcolsep}{0.3em} 
\renewcommand{\arraystretch}{1.1}
\begin{tabular}{|l|l|l|l|l|l|l|l|}
\hline
 \thead{Model} &  \thead{$|\params|$}  & \thead{$|Ap(t)|$} & \thead{\# WB states} & \thead{\# SB states} & \thead{Time} \\
\hline
 \multirow{3}{*}{Cell-Fate} & 1 & 1 & 258,000 -- 491,184 & 32 -- 352 & 4.4 -- 9.19 s \\ 
                            & 8 & 1 -- 4 & 258,000 -- 491,464 & 21 -- 79 & 4.63 -- 13.42 s \\ 
                            & 20 & 7 -- 56 & 258,048 -- 491,520 & 1632 -- 262,144 & 5.82 -- 26.59 s \\ 
\hline
 \multirow{4}{*}{Myeloid}   & 1 & 1 & 128 -- 1,152 & 64 -- 384 & 8 -- 30 ms \\ 
                            & 32 & 63 -- 2,052 & 224 -- 1,984 & 64 -- 1,472 & 14 -- 214 ms \\ 
                            & 70 & \num{5.9e4} -- \num{1.8e7} & 1,512 -- 2,048 & 256 -- 2,048 & 147 -- 1717 s \\ 
                            & 94 & \num{3.4e6} -- \num{3.7e9} & 2,008 -- 2,048 & 1,024 -- 2,048 & 0.6 -- 15.38 s \\ 
\hline
\end{tabular}
\label{table:results}
\end{table}

Next, we ``virtually'' compare our approach to the na\"ive parameter-scan approach. In~\cite{Baudin2019},
a strong basin of (non-parametrised) asynchronous BNs is computed
using a block decomposition method with 4 ms needed to finish the computation for the \emph{myeloid model}. Even if the reported HW was
slower than in our case and we assume that the strong basin computation for one parametrisation would last only 1 ms, the fully parametrised myeloid model contains an attractor which is present in \num{3.7e9} parametrisations. Therefore, the expected time for computing a strong basin for all parametrisations with 32-fold parameterisation would last more than a day compared to less than 27 seconds achieved using our parameter-based semi-symbolic approach. 

We evaluate the scalability of our approach on a fully parametrised myeloid model. The results are shown in Table~\ref{table:scalability}. The computation was restricted to the specified amount of CPUs. The final speed-up achieved on our machine, when using 32 CPUs compared to a non-parallel CPU usage was about 10-fold.

\enlargethispage*{5mm}
\begin{table}[h]
\caption{Scalability of strong basin computation. The strong basins are computed on attractors of myeloid model .}
\centering
\setlength{\tabcolsep}{0.3em} 
\renewcommand{\arraystretch}{1.1}
\begin{tabular}{|c|l|l|l|l|l|l|l|}
\hline
 \thead{\# CPUs} &  \thead{Attractor 1}  & \thead{Attractor 2} & \thead{Attractor 3} & \thead{Attractor 4} & \thead{Attractor 5} & \thead{Attractor 6} \\
\hline
  \textbf{1} & 6.13 s & 99.31 s & 71.32 s & 130.17 s & 45.84 s & 136.65 s \\
\hline
\textbf{2} & 3.34 s & 54.32 s & 38.87 s & 71.31 s & 24.95 s & 74.29 s \\
\hline
\textbf{4} & 1.86 s & 31.3 s & 21.83 s & 40.31 s & 13.71 s & 42.26 s \\
\hline
\textbf{8} & 1.11 s & 19.34 s & 13.31 s & 24.68 s & 8.4 s & 26.49 s \\
\hline
\textbf{16} & 0.87 s & 14.03 s & 9.32 s & 17.56 s & 5.77 s & 18.86 s \\
\hline
\textbf{32} & 0.6 s & 10.98 s & 6.87 s & 13.22 s & 4.54 s & 15.38 s \\
\hline
\end{tabular}
\label{table:scalability}
\end{table}

Now let us have a look at an example of how the results might be interpreted and an suitable control selected. Suppose that we want to
reprogram an erythrocyte cell (phenotype having factors EKLF=1 and
GATA-2=0) into a monocyte cell (phenotype having factors cJun=1 and
EgrNab=1) of the myeloid model. First, we obtain a strong basin of the
monocyte attractor having 1472 states yielding us 1472 possible
control sets. We can observe, that trivial; i.e., the most robust
control strategy (setting variables so that we reach the attractor
right after applying the control) has size 8. We can discard all controls with bigger size 8 and bigger as they are less optimal in all aspects than the trivial control. In our case, there are 1259 controls with smaller size than the trivial one. 

Resulting smallest control sets have the size 1. However, the best
robustness among these controls is 46\%. Therefore, it is quite likely that it will not work in the practice; for a real cell. If we allow the control to have the size 2, we can use the control with 76.8\% robustness. Further increasing the size provides control with the size
3 and the robustness 92\% and so is highly likely to reprogramm the cell's phenotype. To achieve only slightly more robustness (93.8\%) we need to use a control of the size 5. In this highly parametrised model, there is no control with 100\% robustness being smaller than the trivial one. It is not a rule of the thumb that with bigger size of control we obtain better robustness, for example, the control with the size 11 (the only one, with all variables, reversed compared to the original one) has the robustness of only 50\%. 

It can be seen that the unknown properties of \PBN{} make selection of one particular ``best`` control complicated. That is why a careful in vitro experimentation is needed to verify the correctness of the control in the reality. Nonetheless, the obtained control set still can help and highly reduce the exponential number of potential transcription factors combinations.

\section{Conclusion}
\enlargethispage*{6mm}
We have introduced the control problem for parametrised
Boolean networks and we proposed a algorithm for solving  the
source-target variant of this problem using one-step
perturbations. The core procedure of the algorithm is a fixed-point
computation of the parametrised strong basin of the given target. The
method we proposed is semi-symbolic -- it relies on a unique
integration of symbolic (based on BDD representation) and explicit (based on
enumerative model
checking) formal methods. We
demonstrated  that our approach is capable to control highly
parametrised models in seconds. Owing to the doubly-exponential
explosion of the number of possible parametrisations, such a result
cannot be achieved with the na\"ive parameter scan approach. 

In the future work, we would like to scale up the algorithm for searching the strong basin, i.e., by using some symbolic approach for exploring \PBN{'s} state transition graph. Moreover, we would like to consider other variants of \PBN{} control problems based on variants of non-parametrised control problems (see Subsection \ref{ss:BNControl}). Last but not least, we would like to incorporate the developed methods into an existing \PBN{} toolkit -- the AEON tool \cite{aeon}.

\bibliographystyle{unsrt}

\bibliography{references}

\end{document}